\documentclass[11pt]{article}
\usepackage{float}
\usepackage{graphicx}
\usepackage{amsmath}

\def \lket {|}
\def \rket {\rangle}
\def \lbra {\langle}
\def \rbra {|}
\def \Erf{\textnormal{Erf}}
\def \Erfc{\textnormal{Erfc}}

\def\bbbr{R}
\newcommand{\ket}[1]{\lket #1\rket}
\newcommand{\bra}[1]{\lbra #1\rbra}
\newcommand{\comment}[1]{}

\floatstyle{boxed}
\newfloat{Algorithm}{hbt}{lop}
\floatname{Algorithm}{Algorithm}

\newtheorem{Theorem}{Theorem}
\newtheorem{Lemma}{Lemma}
\newtheorem{Corollary}{Corollary}
\newtheorem{Claim}{Claim}
\newcommand{\proof}{\noindent {\bf Proof: }}
\newcommand{\qed}{\nobreak \ifvmode \relax \else
      \ifdim\lastskip<1.5em \hskip-\lastskip
      \hskip1.5em plus0em minus0.5em \fi \nobreak
      \vrule height0.75em width0.5em depth0.25em\fi}

\begin{document}

\title{Quantum strategies are better than classical in almost any XOR game\thanks{Supported by ESF project 1DP/1.1.1.2.0/09/APIA/VIAA/044,
FP7 Marie Curie International Reintegration
Grant PIRG02-GA-2007-224886 and FP7 FET-Open project QCS.}}
\author{
Andris Ambainis,
Art\=urs Ba\v ckurs,
Kaspars Balodis, \\
Dmitrijs Krav\v cenko, 
Raitis Ozols,
Juris Smotrovs,
Madars Virza\\ \\
Faculty of Computing, University of Latvia, \\
Rai\c na bulv. 19, R\=\i ga, LV-1586, Latvia
}

\date{}

\maketitle

\abstract{
We initiate a study of random instances of nonlocal games.
We show that quantum strategies are better than classical for
almost any 2-player XOR game.
More precisely, for large $n$, the entangled value of a random 2-player XOR game with $n$ 
questions to every player is at least 1.21... times the classical value, for $1-o(1)$ fraction
of all 2-player XOR games.
}

\section{Introduction}

Quantum mechanics is strikingly different from classical physics. 
In the area of information processing, this difference can be seen through quantum
algorithms which can be exponentially faster than 
conventional algorithms \cite{Simon,Shor} and 
through quantum cryptography which offers degree of security 
that is impossible classically \cite{BB84}.

Another information-theoretic way of seeing the difference between 
quantum mechanics and the classical world is through non-local games. 
An example of a non-local game is the CHSH (Clauser-Horne-Shimony-Holt) game \cite{CHSH}. 
This is a game played by two players against a referee. 
The two players cannot communicate but can 
share common randomness or a common quantum state that is prepared before the beginning
of the game. The referee sends an independent uniformly random bit to each of the two players. 
Each player responds by sending one bit back to the referee. 
Players win if $x\oplus y = i\wedge j$ where $i, j$ are the bits that the referee sent to the 
player and $x, y$ are players' responses. 
The maximum winning probability that can be achieved is 0.75 classically and 
$\frac{1}{2}+\frac{1}{2\sqrt{2}} = 0.85...$
quantumly. 

There are several reasons why non-local games are interesting. First, CHSH game provides
a very simple example to test the validity of quantum mechanics. 
If we have implemented the referee and the two players $A$, $B$ 
by devices so that there is no communication possible between $A$ and $B$ and 
we observe the winning probability of 0.85..., 
there is no classical explanation possible. 
Second, non-local games 
have been used in device-independent cryptography \cite{Device1,Device2}.

Some non-local games show big gaps between the classical and the 
quantum winning probabilities. For example, Buhrman et al. \cite{Buhrman}
construct a 2-player quantum game where the referee and the players send values 
$x, y, i, j\in\{1, \ldots, n\}$ and the classical winning probability
is $\frac{1}{2}+\Theta(\frac{1}{\sqrt{n}})$ while the quantum winning 
probability is 1. In contrast, Almeida et al. \cite{Almeida} construct a
non-trivial example of a game in which quantum strategies provide no advantage at all.

Which of those is the typical behaviour? 
In this paper, we study this question by looking at random instances
of non-local games.

More specifically, we study two-party XOR games. 
This is a subclass of non-local games with 2 players, 
where the referee randomly chooses inputs 
$i\in\{1, 2, \ldots, n\}$, $j\in\{1, 2, \ldots, k\}$ and sends them to the players.
The players reply by sending bits $x$ and $y$.
The rules of the game are specified by an $n\times k$ matrix $A$
whose entries are $+1$ and $-1$.
To win, the players must produce $x$ and $y$ with $x=y$ if $A_{ij}=1$
and $x$ and $y$ with $x\neq y$ if $A_{ij}=-1$.

\comment{
We choose to study XOR games because there is a well-developed set of
methods for studying them \cite{T,Cleve,Wehner}. The questions that we study 
are quite difficult mathematically }

We consider the case when the matrix $A$ that specifies the rules of the game
is chosen randomly against all $\pm 1$-valued $n\times k$ matrices $A$.
For the case when $n=k$, we show that
\begin{itemize}
\item
The maximum winning probability $p_{q}$ that can be achieved by a quantum strategy is
$\frac{1}{2}+\frac{1\pm o(1)}{\sqrt{n}}$ with a probability $1-o(1)$;
\item
The maximum winning probability $p_{cl}$ that can be achieved by a classical strategy
satisfies 
\[ \frac{1}{2} + \frac{0.6394...-o(1)}{\sqrt{n}} \leq p_{cl} \leq
\frac{1}{2} + \frac{0.8325...+o(1)}{\sqrt{n}}
\] with a probability $1-o(1)$.
\end{itemize}
In the literature on non-local games, one typically studies the difference
between the winning probability $p_q$ ($p_{cl}$) and the
losing probability $1-p_q$ ($1-p_{cl}$): 
$\Delta_q=2p_q-1$ ($\Delta_{cl}=2p_{cl}-1$).
The advantage of quantum strategies is then
evaluated by the ratio $\frac{\Delta_{q}}{\Delta_{cl}}$.
For random XOR games, our results imply that
\[ 1.2011... < \frac{\Delta_{q}}{\Delta_{cl}} < 1.5638... \]
for almost all games.
Our computer experiments suggest that, for large $n$, $\frac{\Delta_q}{\Delta_{cl}} \approx 1.305...$.
For comparison, the biggest advantage that can be achieved in any 2-player XOR game is 
equal to Grothendieck's constant $K_G$ \cite{G} about which we know that \cite{Krivine,Reeds,Braverman}
\[ 1.67696.... \leq K_G \leq 1.7822139781... \]
Thus, the quantum advantage in random XOR games is 
comparable to the maximum possible advantage for
this class of non-local games.

We find this result quite surprising. Quantum-over-classical advantage usually makes use of a structure that is present in the computational problem (such as the algebraic structure that enables Shor's quantum algorithm
for factoring \cite{Shor}). Such structure is normally not present in random computational problems. 

The methods that we use to prove our results are also quite interesting. 
The upper bounds are easy in both classical and quantum case but both
lower bounds are fairly sophisticated. The lower bound on the classical 
value of random XOR games requires a subtle argument that reduces lower-bounding
the classical value to analyzing a certain random walk. The lower bound for
the entangled value requires proving a new version of 
Mar\v cenko-Pastur law \cite{MP} for random matrices. 

\comment{
For example, computing a random Boolean circuit requires $(1+o(1)) \frac{2^n}{n}$ 2-bit (or 2-qubit) gates 
both classically \cite{Wegener} and quantumly \cite{Knill}.
In the query model, computing a random Boolean function
\cite{BW} requires $n$ queries classically and between
$\frac{n}{4}-o(n)$ and $\frac{n}{2}+o(n)$ queries quantumly \cite{A,Dam}. 
At the same time, much bigger speedup can be achieved for non-random problems.
For example, Grover's quantum search \cite{Grover}
algorithm provides a quadratic speedup over the best classical algorithm.

In our case, the situation is very different. Both 
quantum advantage achievable for random XOR games and the biggest possible quantum advantage
are constant factors of similar magnitudes.
}

{\bf Related work.}
Junge and Palazuelos \cite{JP} have constructed non-local games with a big gap between the 
quantum (entangled) value and the classical value, via randomized constructions. 
The difference between this paper and \cite{JP} is as follows. 
The goal of \cite{JP} was to construct a big gap between the entangled value 
and the classical value of a non-local game and the probability distribution on non-local 
games and inputs was chosen so that this goal would be achieved.

Our goal is to study the behaviour of non-local games in the case when the conditions are 
random. We therefore choose a natural probability distribution on non-local games
(without the goal of optimizing the quantum advantage) and study it. The surprising fact is 
that a substantial quantum advantage still exists in such setting.

\section{Technical preliminaries}

We use $[n]$ to denote the set $\{1, 2, \ldots, n\}$.

In a 2-player XOR game, we have two players $A$ and $B$ playing against a referee.
Players $A$ and $B$ cannot communicate but can 
share common random bits (in the classical case) or
an entangled quantum state (in the quantum case). The referee randomly chooses values $i\in\{1, \ldots, n\}$ and 
$j\in\{1, \ldots, n\}$ and sends them to $A$ and $B$, respectively. Players $A$ and $B$ 
respond by sending answers $x\in\{0, 1\}$ and $y\in\{0, 1\}$ to the referee. 

Players win if answers $x$ and $y$ satisfy some winning condition $P(i, j, x, y)$. 
For XOR games, the condition may only depend on the parity $x\oplus y$ of players' responses.
Then, it can be written as $P(i, j, x\oplus y)$.

For this paper, we also assume that, for any $i, j$, exactly one of $P(i, j, 0)$
and $P(i, j, 1)$ is true.
Then, we can describe a game 
by an $n\times n$ matrix $(A_{ij})_{i, j=1}^n$ where $A_{ij}=1$ 
means that, given $i$ and $j$, players must output $x, y$ with $x\oplus y=0$ (equivalently, $x=y$) and $A_{ij}=-1$ 
means that players must output $x, y$ with $x\oplus y=1$
(equivalently, $x\neq y$). 

Let $p_{S, win}$ be the probability that the players win if they use a strategy $S$
and $p_{S, los}=1-p_{S, win}$ be the probability that they lose.
We will be interested in the difference $\Delta_S=p_{S, win}-p_{S, los}$ between
the winning and the losing probabilities. 
The {\em classical value} of a game, $\Delta_{cl}$, is the maximum of $\Delta_S$ over
all classical strategies $S$.
The {\em entangled value} of a game, $\Delta_{q}$, is the maximum of $\Delta_S$ over
all quantum strategies $S$.

Let $p_{ij}$ be the probability that the referee sends question $i$ to player $A$ and
question $j$ to player $B$. Then \cite[section 5.3]{Cleve}, the classical value of the game is equal to 
\begin{equation}
\label{eq:clval} 
\Delta_{cl}=\max_{u_1, \ldots, u_n \in\{-1, 1\}} \max_{v_1, \ldots, v_n \in\{-1, 1\}}
\sum_{i, j=1}^n p_{ij} A_{ij} u_i v_j .
\end{equation}
\comment{To see that, we first observe that players win if and only if $A_{ij} (-1)^{x\oplus y}=1$
and lose if and only if $A_{ij} (-1)^{x\oplus y} = -1$. Let 
$u_i = (-1)^{x_i}$ and $v_j=(-1)^{y_j}$
where $x_i$ ($y_j$) is the value that Player $A$ ($B$) sends to the referee if
he receives question $i$ ($j$).
Then, $A_{ij} (-1)^{x_i\oplus y_i}= A_{ij} u_i v_j$.
Hence, 
\[ p_{S, win} = \sum_{i, j: A_{ij} u_i v_j =1} p_{ij} A_{ij} u_i v_j 
\mbox{~and~} 
p_{S, los} = \sum_{i, j: A_{ij} u_i v_j =-1} - p_{ij} A_{ij} u_i v_j .\]
This means that $\Delta_S = p_{S, win}-p_{S, los} = \sum_{ij} A_{ij} u_i v_j$. 
Taking maximum over all possible strategies $u_1, \ldots, u_n$ and $v_1, \ldots, v_n$
implies (\ref{eq:clval}).}

In the quantum case, Tsirelson's theorem \cite{T} implies that
\begin{equation}
\label{eq:qvalue} 
\Delta_{q}= \max_{u_i: \|u_i\|=1} \max_{v_j: \|v_j\|=1}
\sum_{i, j=1}^n p_{ij} A_{ij} (u_i, v_j) 
\end{equation}
where the maximization is over all tuples of unit-length
vectors $u_1, \ldots, u_n\in \bbbr^d$,
$v_1, \ldots, v_n\in\bbbr^d$ (in an arbitrary number of dimensions $d$).

We will assume that the probability distribution on the referee's questions $i, j$
is uniform: $p_{ij}=\frac{1}{n^2}$ and study $\Delta_{cl}$ and $\Delta_q$ for
the case when $A$ is a random Bernoulli matrix (i.e., each entry $A_{ij}$ is $+1$ 
with probability $1/2$ and $-1$ with probability $1/2$, independently of other entries).

Other probability distributions on referee's questions can be considered, as well. 
For example, one could choose $y_{ij}$ to be 
normally distributed random variables with mean 0 and 
variance 1 and take $p_{ij}=\frac{|y_{ij}|}{\sum_{i, j=1}^n |y_{ij}|}$. 
Or, more generally, one could start with $y_{ij}$ being i.i.d. random variables from
some arbitrary distribution $D$ and define $p_{ij}$ in a similar way.

Most of our results are still true in this more general setting (with mild assumptions on
the probability distribution $D$). Namely, Theorem \ref{thm:q} and the upper bound part
of Theorem \ref{thm:cl} remain unchanged. The only exception is the 
lower bound part of Theorem \ref{thm:cl} which
relies on the fact that the probability distribution $p_{ij}$ is uniform. 
It might be possible to generalize our lower bound proof 
to other distributions $D$ but the exact constant in such generalization of
our lower bound could depend on the probability distribution $D$.

\section{Quantum upper and lower bound}

\begin{Theorem}
\label{thm:q}
For a random 2-player XOR game with $n$ inputs for each player, 
\[ \Delta_{q} = \frac{2\pm o(1)}{\sqrt{n}} \]
with probability $1-o(1)$. 
\end{Theorem}

\proof
Because of (\ref{eq:qvalue}), proving our theorem is equivalent to showing that
\[ \max_{\|u_i\|=\|v_j\|=1} 
\sum_{i=1}^n \sum_{j=1}^n A_{ij} (u_i, v_j) = (2\pm o(1)) n^{3/2} \]
holds with probability $1-o(1)$. 

For the upper bound,  we rewrite this expression as follows. Let $u$ be a vector obtained by concatenating all vectors $u_i$ and 
$v$ be a vector obtained by concatenating all $v_j$. Since $\|u_i\|=\|v_j\|=1$, we have $\|u\|=\|v\|=\sqrt{n}$.
We have
\[ \sum_{i=1}^n \sum_{j=1}^n A_{ij} (u_i, v_j) = (u, (A\otimes I) v) \leq \|u\| \cdot \|A \otimes I\| \cdot \|v\| \leq \|A\| n.\]
By known results on norms of random matrices \cite{Tao}, 
$\|A\|=(2+o(1))\sqrt{n}$ with a high probability.

For the lower bound, we note that 
\[ \max_{\|u_i\|=\|v_j\|=1} \sum_{i=1}^n \sum_{j=1}^n a_{ij} (u_i, v_j)
= \max_{\|u_i\|\leq 1, \|v_j\|\leq 1} \sum_{i=1}^n \sum_{j=1}^n a_{ij} (u_i, v_j) .\]
We have

\begin{Theorem}[Mar\v cenko-Pastur law, \cite{MP}]
Let $A$ be a $n\times n$ random matrix whose entries $A_{ij}$ are independent random
variables with mean 0 and variance 1.
Let $C\in[0, 2]$. 
With probability $1-o(1)$, the number of 
singular values $\lambda$ of $A$ that satisfy $\lambda\geq C \sqrt{n}$
is $(f(C)-o(1))n$ where 
\[ f(C)=\frac{1}{2\pi} \int_{x=C^2}^4 \sqrt{\frac{4}{x}-1} dx .\] 
\end{Theorem}

Let $\lambda_1, \ldots, \lambda_m$ be the singular values of $A$ that satisfy 
$\lambda_i \geq (2-\epsilon) \sqrt{n}$. With high probability, we have
$m\in [(f(2-\epsilon)-o(1)) n, (f(2-\epsilon)+o(1))n]$. We now assume that 
this is the case.

Let $l_i$ and $r_i$ be the corresponding 
left and right singular vectors: $A r_i = \lambda_i l_i$.
(Here, we choose $l_i$ and $r_i$ so that $\| l_i\|=\|r_i\|=1$ for all $i$.) 
Let $l_{ij}$ and $r_{ij}$ be the components of $l_i$ and $r_i$: 
$l_i = (l_{ij})_{j=1}^n$ and $r_i = (r_{ij})_{j=1}^n$. 

We define $u_j$ and $v_j$ in a following way:
\[ u_j = ( l_{ij})_{i=1}^m, \mbox{~~~~}  
v_j = ( r_{ij})_{i=1}^m .\]
We have 
\[ \sum_{i=1}^n \sum_{j=1}^n a_{ij} (u_i, v_j) = 
\sum_{i=1}^n \sum_{j=1}^n \sum_{k=1}^m a_{ij} 
l_{ki} r_{kj} \]
\begin{equation}
\label{eq:sum1} 
= \sum_{k=1}^m (l_k, A r_k) = \sum_{k=1}^m \lambda_k \geq (2-\epsilon)m \sqrt{n} . 
\end{equation}
Since $\|l_i\|=\|r_i\|=1$ and the vectors $u_i$ and $v_j$ are obtained by rearranging 
the entries of $l_i$ and $r_i$, we have
\[ \sum_{i=1}^n \| u_i\|^2 = \sum_{i=1}^n \|l_i \|^2 = m \]
and, similarly, $\sum_i \| v_i\|^2 = m$.  
If $u_i$ and $v_i$ all were of the same length, we would have 
$\|u_i\|^2=\|v_i\|^2 = \frac{m}{n}$. Then, replacing $u_i$ and $v_i$ by
$u'_i = \frac{u_i}{\|u_i\|}$ and $v'_i = \frac{v_i}{\|v_i\|}$ would increase
each vector $\sqrt{\frac{n}{m}}$ times and result in 
\[ \sum_{i=1}^n \sum_{j=1}^n a_{ij} (u'_i, v'_j) \geq (2-\epsilon) n^{3/2} .\]

To deal with the general case, we will show that almost all 
$u_i$ and $v_i$ are of roughly the same length. 
Then, a similar argument will be used.
The key to our proof is a new modification of Mar\v cenko-Pastur law.

\begin{Theorem}[Modified Mar\v cenko-Pastur law]
\label{thm:modified}
Let $A$ be an $n\times n$ random matrix whose entries $A_{ij}$ are independent random
variables with mean 0 and variance 1.
Let $C\in[0, 2]$.
Let $e_i$ be the $i^{\rm th}$ vector of the standard basis. 
Let $P_C$ be the projector on the subspace spanned by the right singular vectors
with singular values at least $C\sqrt{n}$. 
Then, 
\[ Pr\left[ \left| \|P_C e_i\|^2 - f(C) \right| > \epsilon \right] = 
O\left(\frac{1}{n} \right) \]
with the big-O constant depending on $C$ and $\epsilon$. 
\end{Theorem}

The same result also holds for the left singular vectors.

\proof
In appendix \ref{sec:modified}.
\qed

We now complete the proof, assuming the modified Mar\v cenko-Pastur law.
Since $P_C$ is spanned by the right singular vectors $r_1, \ldots, r_m$, we have
\begin{equation}
\label{eq:proj} 
\| P_C e_i \|^2 = \sum_{j=1}^m (r_j, e_i)^2 = \sum_{j=1}^m r_{ji}^2 = \| v_i \|^2 .
\end{equation}
Therefore, the modified Mar\v cenko-Pastur law means that 
\[ Pr[ \|v_i\|^2 > f(2-\epsilon) + \delta] =  O\left(\frac{1}{n} \right) .\]
Thus, the expected number of $i\in\{1, \ldots, n\}$ for which 
$\|v_i\|^2 > f(2-\epsilon) + \delta$ is $O(1)$. We now apply the following 
transformations to vectors $v_i$:
\begin{enumerate}
\item
For each $v_i$ with $\|v_i\|^2 > f(2-\epsilon) + \delta$
(or $u_i$ with $\|u_i\|^2 > f(2-\epsilon) + \delta$), we replace it by the zero
vector $\overrightarrow{0}$;
\item
We replace each $v_i$ by \[ v'_i=\frac{v_i}{\sqrt{f(2-\epsilon)+\delta}} \] and
similarly for $u_i$.
\end{enumerate}
After the first step $\|v_i\|^2 \leq f(2-\epsilon) + \delta$ for all $i$.
Hence, after the second step, $\|v'_i\|^2 \leq 1$ for all $i$.

We now bound the effect of those two steps on the sum
\[ \sum_{i=1}^n \sum_{j=1}^n a_{ij} (u_i, v_j) .\]
Because of (\ref{eq:sum1}), the initial value of this sum is at least
\begin{equation}
\label{eq:sum2} 
(2-\epsilon)m \sqrt{n} \geq (2-\epsilon) (f(2-\epsilon)-o(1)) n^{3/2} .
\end{equation}
Because of (\ref{eq:proj}), $\|v_j \|^2 = \|P_C e_j \|^2 \leq \|e_j \|^2 =1$.
Similarly, $\|u_i\|^2 \leq 1$. Hence, $|(u_i, v_j)|\leq 1$ and replacing one 
$v_j$ (or $u_i$) by 0 changes the sum by at most $\sum_{i=1}^n |a_{ij}|=n$.
Replacing $O(1)$ $v_j$'s (or $u_i$'s) changes it by $O(n)$.
Since the sum (\ref{eq:sum2}) is of the order $\Theta(n^{3/2})$, this 
is a lower order change.

Replacing $v_i$'s by $v'_i$'s (and $u_i$'s by similarly defined $u'_i$'s)
increases each inner product $(u_i, v_j)$ $\frac{1}{f(2-\epsilon)+\delta}$ times 
and achieves
\[ \sum_{i=1}^n \sum_{j=1}^n a_{ij} (u'_i, v'_j) \geq  
\frac{(2-\epsilon) (f(2-\epsilon)-o(1))}{f(2-\epsilon)+\delta} n^{3/2} .\]
Since this can be achieved for any fixed $\epsilon>0$ and $\delta>0$, we get that
\[ \max_{\|u'_i\|\leq 1, \|v'_j\|\leq 1} 
\sum_{i=1}^n \sum_{j=1}^n a_{ij} (u'_i, v'_j) \geq  (2-o(1)) n^{3/2} .\]
\qed

\section{Classical upper and lower bound}

In the classical case, we have to estimate
\begin{equation}
\label{eq:max} 
\max_{u_1, \ldots, u_n \in\{-1, 1\}} \max_{v_1, \ldots, v_n \in\{-1, 1\}}
\sum_{i, j=1}^n A_{ij} u_i v_j .
\end{equation}
There are several ways how one can interpret this expression.
First, (\ref{eq:max}) is equal to 
the $l_{\infty}\rightarrow l_1$ norm of $A$
(denoted $\|A\|_{\infty\rightarrow 1}$).
It is known that, for a random matrix $A$,
$\|A\|_{\infty\rightarrow 1}=\Theta(n\sqrt{n})$ (e. g., from \cite{Montero}).
but the exact constant under $\Theta$ is not known.

Gittens and Tropp \cite{Tropp} show that, if $A$ is a matrix whose entries
are i.i.d. random variables, then
\[ \|A\|_{\infty\rightarrow 1} \leq 2 E \left( \|A\|_{col}+\|A^T\|_{col} \right) \]
where $\|A\|_{col}$ denotes the sum of the $l_2$ norms of the columns of $A$.
For our case, this gives an upper bound of 
$\|A\|_{\infty\rightarrow 1}\leq 4 n \sqrt{n}$ which is substantially weaker than
our Theorem \ref{thm:cl} below (and would be insufficient to show a gap between 
classical and entangled values for random XOR games).

In the context of statistical physics, there has been substantial work on determining
the order of 
\begin{equation}
\label{eq:parisi}
 \max_{u_1, \ldots, u_n \in\{-1, 1\}} 
\sum_{i, j=1}^n A_{ij} u_i u_j 
\end{equation}
when $A_{ij}$ is a symmetric Gaussian matrix (each $A_{ij}=A_{ji}$ is an independent
Gaussian random variable with mean 0 and variance 1). It is known that (\ref{eq:parisi}) is equal to
$(1.527...+o(1))n^{3/2}$ with probability $1-o(1)$.
This was first discovered in \cite{SK,Parisi} and rigorously proven by
Talagrand \cite{Talagrand}. 

The quantities (\ref{eq:max}) and (\ref{eq:parisi})
are of similar flavour but are not identical and the work on (\ref{eq:parisi}) does not
directly imply anything about our problem. b

One can also interpret (\ref{eq:max}) combinatorially, as a problem
of ``unbalancing lights" \cite{Unbalancing}. In this interpretation,
$n\times n$ matrix represents an array of lights, with each light being ``on"
($A_{ij}=1$) or ``off" ($A_{ij}=-1$). We are allowed to choose a row or a column
and switch all lights in this row or column. The task is to maximize the difference
between the number of lights that is on and the number of lights that is off.
It is known that for any $n\times n$ matrix $A$ with $\pm 1$ entries, (\ref{eq:max})
is at least $\sqrt{\frac{2}{\pi}} n^{3/2}$ \cite[p.19]{Unbalancing}. 
We were not able to find any work on evaluating (\ref{eq:max}) for a random matrix $A$
in this context.

\begin{Theorem}
\label{thm:cl}
For a random 2-player XOR game, its classical value $\Delta_{cl}$ satisfies
\[ \frac{1.2789...}{\sqrt{n}} \leq 
\Delta_{cl} \leq \frac{2\sqrt{\ln{2}}+o(1)}{\sqrt{n}} =  \frac{1.6651...+o(1)}{\sqrt{n}} 
\]
with probability $1-o(1)$.
\end{Theorem}

This is equivalent to
\[ 1.2789... n^{3/2} \leq \|A\|_{\infty\rightarrow 1} \leq 1.6651...n^{3/2} \]
for a Bernoulli random matrix $A$.

\begin{figure}[h]
\centering
\includegraphics[scale=0.6]{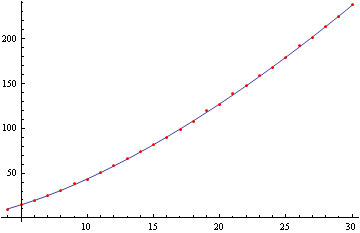}
\caption{$\|A\|_{\infty\rightarrow 1}$, for random $n\times n$ matrices $A$}
\label{fig:1}
\end{figure}

In computer experiments, the ratio $\frac{\|A\|_{\infty\rightarrow 1}}{n^{3/2}}$
grows with $n$ and reaches 1.4519... for $n=26$.  By fitting a formula $an^{3/2}+bn$
where the leading term is of the order $n^{3/2}$ and the largest 
correction term is of the order $n$ to the data, we obtained that
\[ \|A\|_{\infty\rightarrow 1} \approx 1.53274... n^{3/2}-0.472806... n .\]
Figure \ref{fig:1} shows the fit. Curiously, the constant in front of $n^{3/2}$ is very close
to the constant 1.527... for the sum (\ref{eq:parisi}). 
We are not sure whether this is a coincidence or there is some connection between the 
asymptotic behaviour of the two sums.

\proof
The upper bound follows straightforwardly from Chernoff bounds. We use the following form
of Chernoff inequality:

\begin{Theorem}
\cite[p.263]{Unbalancing}
\label{thm:chernoff}
Let $X_1, \ldots, X_n$ be independent random variables with $Pr[X_i=1]=Pr[X_i=-1]=\frac{1}{2}$
and let $X=X_1+\ldots+X_n$.
Then,
\[ Pr [ X \geq a] < e^{-\frac{a^2}{2 n}} .\] 
\end{Theorem}
 
Let $x_1, \ldots, x_n\in\{ -1, 1\}$ and $y_1, \ldots, y_n\in\{ -1, 1\}$ be arbitrary.
If $A_{ij}\in\{-1, 1\}$ are uniformly random, then 
$A_{ij}x_i y_j\in\{-1, 1\}$ are also uniformly random.
Hence, $\sum_{i,j}A_{ij} x_i y_j$ is a sum of $n^2$ uniformly random values from $\{-1, 1\}$.
By Theorem \ref{thm:chernoff},
\[ \text{Pr}\left[\sum_{i,j}A_{ij} x_i y_j >Cn^{\frac{3}{2}}\right]
<e^{\frac{-\left(Cn^{\frac{3}{2}}\right)^2}{2n^2}}=\frac{1}{e^{\frac{C^2n}{2}}} .\]
By taking 
$C=2\sqrt{\ln{2}}+2\frac{\sqrt{\ln n}}{\sqrt{n}}$, 
we can ensure that this probability is less than $\frac{1}{2^{2n}n^2}$.
Then, by the union bound, the probability that 
$\sum_{i,j}A_{ij} x_i y_j > Cn^{\frac{3}{2}}$ for some choice of $x_i$'s and $y_j$'s
is less than $2^{2n} \frac{1}{2^{2n}n^2} = \frac{1}{n^2}$.

To prove the lower bound, we first show 

\begin{Lemma}
\label{lem:ce}
Let $A$ be an $n\times n$ random Bernoulli matrix.
Then,
\[ \operatorname{E}_A\left[\max_{u_i, v_j \in \{-1,1\}}\sum_{i,j}u_i v_j A_{ij}\right] \geq (1.2789...-o(1)) n^{3/2} .\]
\end{Lemma}

Let $X=\max_{u_i, v_j \in \{-1,1\}}\sum_{i,j}u_i v_j A_{ij}$. 
By Lemma \ref{lem:ce}, $E[X]\geq (1.2789...-o(1)) n^{3/2}$.
To prove that $X \geq (1.2789...-o(1)) n^{3/2}$ with probability $1-o(1)$,
we show that $X$ is concentrated around $E[X]$.

\begin{Lemma}
\label{lem:azuma}
Let $X=\max_{u_i, v_j \in \{-1,1\}}\sum_{i,j}u_i v_j A_{ij}$ 
for a random $n\times n$ matrix $A$. Then,
\[ Pr\left[ |X - E[X]|  \geq a n \right] < 2 e^{-a^2/8} .\] 
\end{Lemma}

We then apply Lemma \ref{lem:azuma} with $a=\log n$ (or with $a=f(n)$ 
for any other $f(n)$ that has $f(n)\rightarrow \infty$ when $n\rightarrow\infty$ and 
$f(n)=o(\sqrt{n})$) and combine it with Lemma \ref{lem:ce}.

It remains to prove the two lemmas.
\comment{
For the lower bound, the maximum value of the probability of winning 
minus the probability of losing in the classical case of a random XOR game with $n$ possible values of input is given by
\begin{equation}
\label{XORclassical}
{1\over n^2}\operatorname{E}_A\left[\max_{u_i, v_j \in \{-1,1\}}\sum_{i,j}u_i v_j A_{ij}\right]
\end{equation}
where the expected value is taken uniformly over all $n\times n$ matrices $A$ consisting of $-1$ and $1$.
}

\proof [of Lemma \ref{lem:ce}]
Let $A$ be a random $\pm 1$ matrix.
We choose $u_i$ and $v_j$, according to Algorithm \ref{alg:classical}.
\begin{Algorithm}
\begin{enumerate}
\item Set $u_1=1$.
\item For each $k=2, \ldots, n$ do:
\begin{enumerate}
\item For each $j=1, \ldots, n$, compute 
$S_{k-1,j}=\sum_{i=1}^{k-1} A_{ij}u_i$. 
\item
Let $a_k=(Z(S_{k-1,1}),Z(S_{k-1,2}),...,Z(S_{k-1,n}))$ where $Z(x)=1$ if $x>0$, $Z(x)=-1$ if $x<0$ and $Z(x)=1$ or $Z(x)=-1$ with equal probability ${1 \over 2}$ if $x=0$. 
\item
Let $b_k=(A_{k1},A_{k2},...,A_{kn})$.
\item
Let $u_i\in\{+1, -1\}$ be such that $a_i$ and $b_i u_i$ agree in the maximum number of positions. 
\end{enumerate}
\item For each $j=1, \ldots, n$, let $v_j$ be such that $v_j S_{n,j} \geq 0$
where $S_{n, j}= \sum_{i=1}^n A_{ij} u_i$.
\end{enumerate}
\caption{Algorithm for choosing $u_i$ and $v_j$ for a given matrix $A$.}
\label{alg:classical}
\end{Algorithm}

Because of the last step, we get that
\[ \sum_{i=1}^n \sum_{j=1}^n u_i v_j A_{ij} = \sum_{j=1}^n |S_{n, j}| .\]
Each of $S_{n, j}$ is a random variable with an identical distribution. Hence, 
\begin{equation}
\label{eq:expect} 
E \left[ \sum_{i=1}^n \sum_{j=1}^n u_i v_j A_{ij} \right] 
= \sum_{j=1}^n E |S_{n, j}| = n E|S_{n, 1}| .
\end{equation}

We now consider a random walk with a reflecting boundary. 
The random walk starts at position 0. If it is at the position 0, it always moves to the 
position 1. If it is at the position $i>0$, it moves to the position $i+1$
with probability $\frac{1}{2}+\frac{\epsilon}{2}$ and position $i-1$ with probability 
$\frac{1}{2}-\frac{\epsilon}{2}$.
Let $K^{\epsilon}_{i}$ be the position of the walker after $i$ steps.

\begin{Lemma}
\label{cl:relate}
$|S_{n, 1}| = K^{\epsilon}_n$ for some 
$\epsilon=(1+o(1))\sqrt{{2\over \pi n}}$.
\end{Lemma}

\proof
$b_i= ( A_{i1}, \ldots, A_{in})$ is a vector consisting of random $\pm 1$'s that is 
independent of $a_i$.
Hence, the expected number of agreements between $a_i$ and $b_i u_i$ 
is $({1 \over 2}+{\epsilon \over 2}) n$ where 
$\epsilon=(1+o(1))\sqrt{{2\over \pi n}}$ \cite[p.21]{Unbalancing}.
Moreover, the probability of $a_i$ and $b_i u_i$ agreeing in 
location $j$ is the same for all $j$.

Hence, if $|S_{i-1, 1}|>0$, we have $|S_{i, 1}| = |S_{i-1, 1}|+1$ with probability 
$\frac{1}{2}+\frac{\epsilon}{2}$ and $|S_{i, 1}|=|S_{i-1, 1}|-1$ with probability $\frac{1}{2}-\frac{\epsilon}{2}$.
If $|S_{i-1, 1}|=0$, then we always have $|S_{i, 1}|=1$.
\qed

\begin{Lemma}
\label{cl:walk}
For a random walk with a reflecting boundary and $\epsilon=\frac{\alpha}{\sqrt{n}}$,
we have $E[K^{\epsilon}_n] \geq (f(\alpha)-o(1))\sqrt{n}$ where
\[ f(\alpha) = \frac{1}{2}\left(e^{-\frac{\alpha^2}{2}}\sqrt{\frac{2}{\pi}}+\alpha+\left(\frac{1}{\alpha}+\alpha\right)\textnormal{Erf}\left(\frac{\alpha}{\sqrt{2}}\right)\right).\]
\end{Lemma}

\proof
In appendix \ref{sec:walk}.
\qed

By combining (\ref{eq:expect}) and Lemmas \ref{cl:relate} and \ref{cl:walk},
the probability of winning minus the probability of losing 
in the classical case of a random XOR game is at least
$$f\left(\sqrt{\frac{2}{\pi}}\right)\sqrt{n}\cdot n \cdot{1\over n^2}=\frac{2+2 e^{-1/\pi }+(2+\pi ) 
\text{Erf}\left(\frac{1}{\sqrt{\pi }}\right)}{2 \sqrt{2 \pi }}n^{-\frac{1}{2}} $$ $$=1.2789076012442957...n^{-\frac{1}{2}}.$$
\qed

\proof [of Lemma \ref{lem:azuma}]
Let 
\[ f(A_{11}, A_{12}, \ldots, A_{nn}) =
 \max_{u_i, v_j \in \{-1,1\}}\sum_{i,j}u_i v_j A_{ij} .\]
Then, changing one $A_{ij}$ from $+1$ to $-1$ (or from $-1$ to $+1$) changes
$\sum_{i,j}u_i v_j A_{ij}$ by at most 2. 
This means that $f(A_{11}, \ldots, A_{nn})$ changes by at most 2 as well.
In other words, $f$ is 2-Lipschitz. 
By applying Azuma's inequality \cite[p. 303-305]{MU} with $c=2$, $t=n^2$, $\lambda=\frac{a}{2}$,
we get
\[ Pr\left[ \left|f(A_{11}, \ldots, A_{nn}) - E[f(A_{11}, \ldots, A_{nn})]\right|  \geq a n \right] < 2 e^{-a^2/8} .\] 
\qed

\section{Conclusion}

We showed that quantum strategies are better than classical 
for random instances of XOR games. We expect that similar results may be true for 
other classes of non-local games. 

A possible difficulty with proving them is that the mathematical
methods for analyzing other classes of non-local games are much less developed.
There is a well developed mathematical framework for studying XOR games \cite{T,Cleve,Wehner} 
which we used in our paper. But even with that,
some of our proofs were quite involved. Proving a similar result for
a less well-studied class of games would be even more difficult.

\bigskip
{\bf Acknowledgments.} We thank Assaf Naor, Oded Regev and Stanislaw Szarek for useful comments
and references to related work.

\newpage
\begin{appendix}
\section{Modified Mar\v cenko-Pastur law}
\label{sec:modified}

In this section, we prove Theorem \ref{thm:modified}.

Without loss of generality, we assume that $i=1$.
Let $\lambda_1, \ldots, \lambda_n$ be the singular values of $A$ and let $r_i$ be 
the corresponding right singular vectors. Let $\alpha_i = \lbra e_1 | r_i \rket$.

Let $X$ be a random variable that is equal to $\frac{\lambda^2_i}{n}$ with probability $|\alpha_i|^2$.
Let $Y$ be a random variable distributed according to the Mar\v cenko-Pastur law
with parameters $y=1$ and $\sigma=1$ \cite[Section 3.1]{Bai}. 
Then, the probability density function of $Y$ is equal to
\[ P(x) = \frac{1}{2\pi} \sqrt{\frac{4}{x}-1} \]
for $x\in [0, 4]$ and $P(x)=0$ otherwise.
Let $F_X(x)$ and $F_Y(x)$ denote the cumulative distribution functions of $X$ and $Y$ respectively.
We would like to show that $F_X(x)$ and $F_Y(x)$ are close to one another.

We say that a family ${\cal X}$ of random variables is uniformly sub-Gaussian if
there exist constants $c$ abd $C$ such that
\[ Pr[|X| > x] \leq C e^{-c x^2} \] 
for every $X\in{\cal X}$.
From Levy's continuity theorem \cite[p. 99]{Tao}, we have

\begin{Theorem}
\label{thm:moment}
Let $Y$ be a fixed random variable on some interval $[a, b]$ and let ${\cal X}$ be a uniformly sub-Gaussian 
family of random variables. Let $\epsilon>0$ be fixed. There exist $k>0$, $\delta>0$ 
such that if $X\in{\cal X}$ satisfies  
\begin{equation}
\label{eq:mc} | E[X^m] - E[Y^m] | \leq \delta 
\end{equation}
for all $m\in\{1, \ldots, k\}$, then we have 
\[ | F_{X} (x) - F_{Y}(x) | \leq \epsilon \]
for all $x\in[a, b]$.
\end{Theorem}

We select a constant $D$ so that 
\[ Pr[\|A\|\geq D\sqrt{n}] = O\left( \frac{1}{n} \right) .\]
By concentration results for norms of random matrices (Proposition 2.3.10 of \cite{Tao}), 
$D=2+o(1)$ suffices. We define ${\cal X}$ as the family of random variables $X$, for
all $A$ that satisfy $\|A\|\leq D \sqrt{n}$. Since $X=\frac{\lambda^2_i}{n}$ 
and $\lambda_i\leq \|A\|$, we have
$Pr[X>D^2]=0$. Hence, ${\cal X}$ is uniformly sub-Gaussian because 
constants $C, c$ can be chosen so that $C e^{-cx^2}\geq 1$ for 
$x:0\leq x\leq D^2$. 

To show theorem \ref{thm:modified}, it suffices to prove that the random variable $X \in {\cal X}$ satisfies 
the condition (\ref{eq:mc}) for all $m\in\{1, 2, \ldots, k\}$, 
with probability $1-O(\frac{1}{n})$. 
Since $k$ is fixed, this is equivalent to showing 
(\ref{eq:mc}) for each $m\in\{1, \ldots, k\}$ with probability $1-O(\frac{1}{n})$.
For the random variable $Y$, its moments are quite well known:

\begin{Lemma}
\[ E[Y^m] = C_m \] 
where $C_m =\frac{(2m)!}{m! (m+1)!}$ denotes the $m^{\rm th}$ Catalan number.
\end{Lemma}

\proof
From \cite[Lemma 3.1]{Bai}, we have $E[Y^m]=\sum_{i=1}^{m} N(m, i)$ where $N(m, i)$ are Narayana numbers 
$N(m, i)= \frac{1}{i+1} {m \choose i}{m-1 \choose i}$. From \cite[page 237]{Stanley}, we have $\sum_{i=1}^{m} N(m, i)=C_m$.
\qed

We introduce a random variable $M_m$ that is equal to $E[X^m]$ for 
a random matrix $A$. To show that $M_m \in [C_m - \delta, C_m+\delta]$ with a high probability, 
we bound the expectation and the variance of $M_m$ and then use Chebyshev's inequality. We have

\begin{Lemma}
\label{lem:moment}
Let $m>0$ be fixed. Then, for large $n$,
\[ 
E[M_m]= \left(1 + O \left( \frac{1}{n} \right) \right) C_{m} .\]
\end{Lemma}

\proof
Let $B= A^{T} A$. Then, the eigenvalues of $B$ are $\lambda_1^2, \ldots, \lambda_n^2$.
We have
\[ M_m  = \frac{1}{n^m} \sum_{i=1}^n \lambda_i^{2m} |\alpha_i|^2  = \frac{1}{n^m} \bra{e_1} B^m \ket{e_1}.\]
By expanding $\bra{e_1} B^m \ket{e_1}$, we can write
\begin{equation}
\label{eq:mm} 
M_m = \frac{1}{n^m} \sum_{i_1, \ldots, i_{2m-1}=1}^n
a_{i_1, 1} a_{i_{2m-1}, 1} \prod_{j=1}^{m-1} 
a_{i_{2j-1}, i_{2j}} a_{i_{2j+1}, i_{2j}} .
\end{equation}

$E[M_m]$ is equal to $\frac{1}{n^m}$ times the sum of expectations $E[T]$,
\begin{equation}
\label{eq:term} T =  a_{i_1, 1} a_{i_{2m-1}, 1} 
\prod_{j=1}^{m-1} a_{i_{2j-1}, i_{2j}} a_{i_{2j+1}, i_{2j}} .
\end{equation}
If $T$ consists of $a_{j_1, k_1}$, $a_{j_2, k_2}$, $\ldots$, $a_{j_l, k_l}$ occurring 
$c_1, c_2, \ldots, c_l$ times, we have 
\begin{equation}
\label{eq:prod}
E[T] = E[ a_{j_1, k_1}^{c_1}] E[ a_{j_2, k_2}^{c_2}] \ldots E[ a_{j_l, k_l}^{c_l}] .
\end{equation}
If $c_i$ is odd for some $i$, then $a_{j_i, k_i}^{c_i}$ is 1 with probability 1/2 and
-1 with probability 1/2. Hence, $E[a_{j_i, k_i}^{c_i}]=0$.
If all $c_i$ are even, then $a_{j_i, k_i}^{c_i}=1$ and (\ref{eq:prod}) is equal to 1.
Thus, (\ref{eq:mm}) is equal to the number of terms $T$ 
in which each $a_{j_i, k_i}$ occurs an even number of times.
We call such terms {\em good}.

For a term $T$, let
$|T|$ to be the number of different numbers appearing in 
the sequence $1, i_1, i_2, \ldots, i_{2m-1}$.
We say that a good term $T$ is {\em standard} if the sequence 
$1, i_1, i_2, \ldots, i_{2m-1}$ is such that the $i^{\rm th}$ different
number of this sequence is equal to $i$, for every $i\in\{2, \ldots, |T|\}$.
If we have a standard term $T$, we can obtain $(n-1)(n-2)\ldots(n-|T|+1)$ good terms from it,
by replacing all occurrences of $2, \ldots, |T|$ 
with distinct numbers $j_2, \ldots, j_{|T|}\in\{2, \ldots, n\}$. 
Moreover, each of those good terms can be obtained only from one standard term.

Therefore, we have
\[ E[M_m] = \frac{1}{n^m} \sum_{T - \mbox{standard}} (n-1)(n-2)\ldots(n-|T|+1) .\]
Since the sequence $1, i_1, i_2, \ldots, i_{2m-1}$ 
may contain at most $2m$ different numbers, we
have $|T|\leq 2m$.
Since $m$ is a fixed constant, this implies
\[ E[M_m] = \frac{1}{n^m} \sum_{T - \mbox{standard}} \left(1+ O\left(\frac{1}{n} \right) \right) n^{|T|-1} .\]
Let $t$ be the largest possible value for $|T|$. Then,
\begin{equation}
\label{eq:stand} E[M_m] = \frac{1}{n^m} \left(1+ O\left(\frac{1}{n} \right) \right) n^{t-1} \left| \{ T : T \mbox{- standard}, |T|=t \} \right| .
\end{equation}
The contributions from standard terms $T$ with $|T|<t$ can be absorbed 
into the $O(\frac{1}{n})$ factor, since $n^{|T|-1} \leq n^{t-2} = \frac{1}{n} n^{t-1}$ for
each of those terms and the number of standard terms is a constant (it depends only on $t$ and not $n$).

It remains to determine $t$ and to count the standard terms $T$ with $|T|=t$.
We claim that $t=m+1$.

Let $T$ be a standard term.
Consider the graph $G$ with the set of vertices $V=\{1, \ldots, |T|\}$ 
and the set of edges $E$ consisting of all the different pairs $(i, j)$ that appear in the sequence 
$(i_1, 1)$, $(i_1, i_2)$, $(i_3, i_2)$, $\ldots$, $(i_{2m-1}, 1)$.
Since $T$ is good, each such $(i, j)$ appears an even number of times, i.e. at least twice.
Therefore, $|E|\leq \frac{2m}{2}=m$. 

The graph $G$ is connected. Hence, $|V|\leq |E|+1\leq m+1$. Moreover, $|V|=m+1$ is achieved, for example,
by the graph $G$ obtained from a standard term 
\[ (2, 1), (2, 1), (3, 1), (3, 1), \ldots, (m+1, 1), (m+1, 1) \]
in which $i_{2j}=1$ and $i_{2j-1}=j+1$.
It remains to count the standard terms $T$ with $|T|=m+1$.

For each such term, we have $|V|=m+1$ and $|E|=m$. Hence, $G$ must be a tree, with each edge 
occuring exactly twice in the sequence $(i_1, 1)$, $(i_1, i_2)$, $(i_3, i_2)$, $\ldots$, $(i_{2m-1}, 1)$.
We now consider the sequence
\[ (1, i_1), (i_1, i_2), (i_2, i_3), \ldots, (i_{2m-1}, 1) \]
in which all edges are directed so that the sequence is a closed walk in the tree $G$.
The conditions on this sequence that we have are the same as the conditions on 
{\em a non-crossing cycle} in \cite[p. 144]{Tao}.
As shown in \cite[p. 144-145]{Tao}, the number of non-crossing cycles of length $2m$
is equal to the Catalan number $C_{m}$. The lemma follows by substituting this and $t=m+1$ into
equation (\ref{eq:stand}).
\qed

By a similar argument, we have

\begin{Lemma}
\label{lem:moment1}
Let $m>0$ be fixed. Then, for large $n$,
\[ 
E[M^2_m]= \left(1 + O \left( \frac{1}{n} \right) \right) (C_{m})^2 .\]
\end{Lemma}

\proof
The proof is similar to the proof of Lemma \ref{lem:moment}.
We omit it in this version of the paper.
\qed

From Lemmas \ref{lem:moment} and \ref{lem:moment1}, we get the following corollaries.

\begin{Corollary}
\label{cor:1}
Let $m>0$ be fixed. Then, for large $n$,
\[ 
D[M_m]= O \left( \frac{1}{n} \right) C^2_{m} .\]
\end{Corollary}

\proof
Follows from $D[X]=E[X^2]-E^2[X]$, Lemma \ref{lem:moment} and Lemma \ref{lem:moment1}.
\qed

\begin{Corollary}
Let $m>0$ and $\delta>0$ be fixed. Then, for large $n$,
\begin{equation}
\label{eq:x}
Pr \left[ | M_m - C_m | \geq \delta \right] =  O \left( \frac{1}{n} \right)   .
\end{equation}
\end{Corollary}

\proof
Follows from Corollary \ref{cor:1} and Chebyshev inequality.
\qed

This completes the proof of Theorem \ref{thm:modified}.

\section{Analysis of a random walk with a reflecting boundary}
\label{sec:walk}

\subsection{Proof overview}

In this section, we prove Lemma \ref{cl:walk}.
To simplify notation, we write $K_i$ instead of $K^{\epsilon}_i$.
We have
\[ E[K_{i+1}] = Pr[K_i>0] E[K_{i+1} | K_i>0] + Pr[K_i=0] E [K_{i+1} | K_i = 0] .\]
If $K_i=m$ and $m>0$, we have
\[ E[K_{i+1} | K_i=m] = \left( \frac{1}{2}+\frac{\epsilon}{2} \right) (m + 1) + 
\left( \frac{1}{2}-\frac{\epsilon}{2} \right) (m - 1) = m + \epsilon .\]
We also have $E[K_{i+1} | K_i = 0] = 1$. Hence,
\[ E[K_{i+1}] = E[K_i] + \epsilon + (1-\epsilon) Pr[K_i=0] .\]
By induction, this implies 
\[ E[K_n] = \epsilon n+(1-\epsilon)\sum_{i=0}^{n-1}\mbox{Pr}\left[K_i=0\right] .\]
We have $\epsilon n = \alpha \sqrt{n}$. It remains to bound 
$\sum_{i=0}^{n-1}\mbox{Pr}\left[K_i=0\right]$.
The first step is to express this sum in terms of binomial coefficients.

\begin{Claim}
\label{cl:binomial}
$$
\sum_{i=0}^{n-1}\mbox{Pr}\left[K_i=0\right] = 1 +
\sum_{S=0}^{2 \lfloor \frac{n-1}{2} \rfloor -2}
\sum_{A=\max(0,S-\lfloor \frac{n-1}{2}\rfloor + 1)}^{\lfloor \frac{S}{2} \rfloor}
\left(\binom{S}{A}-\binom{S}{A-1}\right)p^{A}(1-p)^{S-A+1} 
$$
where $p=\frac{1}{2}+\frac{\epsilon}{2}$.
\end{Claim}
We can express this sum as a difference of two sums 
$$
\sum_{S=0}^{2 \lfloor \frac{n-1}{2} \rfloor -2}
\sum_{A=\max(0,S-\lfloor \frac{n-1}{2}\rfloor + 1)}^{\lfloor \frac{S}{2} \rfloor}
\left(\binom{S}{A}-\binom{S}{A-1}\right)p^{A}(1-p)^{S-A+1} =
$$
$$
(1-p)\sum_{S=0}^{2 \lfloor \frac{n-1}{2} \rfloor -2}\sum_{A=0}^{\lfloor \frac{S}{2} \rfloor}
\left(\binom{S}{A}-\binom{S}{A-1}\right)p^{A}(1-p)^{S-A}
$$
$$
-(1-p)\sum_{S=\lfloor \frac{n-1}{2} \rfloor}^{2 \lfloor \frac{n-1}{2} \rfloor -2}\sum_{A=0}^{S-\lfloor \frac{n-1}{2}\rfloor}
\left(\binom{S}{A}-\binom{S}{A-1}\right)p^{A}(1-p)^{S-A}.
$$

\begin{Claim}
\label{cl:sum1}
$$
\sum_{S=0}^{2 \lfloor \frac{n-1}{2} \rfloor -2}\sum_{A=0}^{\lfloor \frac{S}{2} \rfloor}
\left(\binom{S}{A}-\binom{S}{A-1}\right)p^{A}(1-p)^{S-A} \geq
$$
$$
(1+o(1)) \left(e^{-\frac{\alpha^2}{2}}\sqrt{\frac{2}{\pi}}-\alpha+\left(\frac{1}{\alpha}+\alpha\right)\textnormal{Erf}\left(\frac{\alpha}{\sqrt{2}}\right)\right) \sqrt{n}
$$
where 
$$
\textnormal{Erf}(z)=\frac{2}{\sqrt{\pi}}\int_0^ze^{-t^2}dt.
$$
is the error function for the normal distribution.
\end{Claim}

\begin{Claim}
\label{cl:sum2}
$$
\sum_{S=\lfloor \frac{n-1}{2} \rfloor}^{2 \lfloor \frac{n-1}{2} \rfloor -2}\sum_{A=0}^{S-\lfloor \frac{n-1}{2}\rfloor}
\left(\binom{S}{A}-\binom{S}{A-1}\right)p^{A}(1-p)^{S-A}<1.
$$
\end{Claim}

Lemma \ref{cl:walk} now follows by combining Claims \ref{cl:binomial}, \ref{cl:sum1}
and \ref{cl:sum2}.

\subsection{Proof of Claim \ref{cl:binomial}}

Let $T(m,t)$ be the number of paths of length $2m$ that start and end at the point $0$, 
increase or decrease the coordinate by $1$ at each step,
never take a negative coordinate and return to the location 0 $t$ times (including the last step).
Then, for any $m>0$ and $t:1\leq t\leq m$, we have \cite[p. 203]{Knuth}, 
\begin{equation}
\label{eq:dyck}
T(m,t)= \frac{t}{m} \binom{2m-t-1}{m-t}.
\end{equation}
We have
$$
\sum_{i=0}^{n-1}\textnormal{Pr}\left[K_i=0\right] =
1+ \sum_{m=1}^{\lfloor \frac{n-1}{2} \rfloor}\textnormal{Pr}\left[K_{2m}=0\right]=
$$
$$
1+ \sum_{m=1}^{\lfloor \frac{n-1}{2} \rfloor}\sum_{t=1}^{m}T(m,t)p^{m-t}(1-p)^m=
$$
\begin{equation}
\label{eq:last}
1 +\sum_{m=1}^{\lfloor \frac{n-1}{2} 
\rfloor}\sum_{t=1}^{m}\binom{2m-t-1}{m-t}\frac{t}{m}p^{m-t}(1-p)^m.
\end{equation}
Here, the first equality follows from the fact that the random walk can return to
location 0 only after an even number of steps. The second equality follows 
by partitioning the paths that return to 0 after $2m$ steps according to the total number of 
times the path returns to 0 (including the final return to 0). 
If the path returns to 0 $t$ times, then it also leaves 0 $t$ times.
Hence, there are $t$ steps in which we move right with probability 1 
(the steps which start at location 0). There are also $m-t$ other
steps when the path moves right (each of those steps is taken with
probability $p$) and $m$ steps when the path moves left (each of those steps is
taken with probability $1-p$).
This means that the probability of each path with $t$ returns is 
$p^{m-t}(1-p)^m$.
The third equality follows from (\ref{eq:dyck}). 

Let $S=2m-t-1$ and $A=m-t$.
We can rewrite the sum (\ref{eq:last}) in the following way:
$$
\sum_{m=1}^{\lfloor \frac{n-1}{2} \rfloor}\sum_{t=1}^{m}\binom{2m-t-1}{m-t}\frac{t}{m}p^{m-t}(1-p)^m =
$$
$$
\sum_{S=0}^{2 \lfloor \frac{n-1}{2} \rfloor -2}\sum_{A=\max(0,S-\lfloor \frac{n-1}{2}\rfloor + 1)}^{\lfloor \frac{S}{2} \rfloor}\binom{S}{A}\frac{S-2A+1}{S-A+1}p^{A}(1-p)^{S-A+1}=
$$
$$
\sum_{S=0}^{2 \lfloor \frac{n-1}{2} \rfloor -2}\sum_{A=\max(0,S-\lfloor \frac{n-1}{2}\rfloor + 1)}^{\lfloor \frac{S}{2} \rfloor} \left(\binom{S}{A}-\binom{S}{A-1}\right)p^{A}(1-p)^{S-A+1}
$$
where the last equality follows from
$$
\binom{S}{A}\frac{S-2A+1}{S-A+1}=\binom{S}{A}-\binom{S}{A-1}.
$$
This completes the proof of the claim.

\subsection{Proof of Claim \ref{cl:sum1}}

We denote $m=2 \lfloor \frac{n-1}{2} \rfloor -2$.
By using the identity
\begin{equation}
\label{eq:middle}\sum_{A=0}^{k}\left(\binom{S}{A}-\binom{S}{A-1}\right)p^A(1-p)^{S-A}=
\end{equation}
$$
\binom{S}{k}p^k(1-p)^{S-k}-\frac{2\epsilon}{1-\epsilon}\sum_{A=0}^{k-1}\binom{S}{A}p^A(1-p)^{S-A} ,
$$
\comment{$$
\binom{S}{X}p^X(1-p)^{S-X}-2\epsilon (1+ o(1)) \sum_{A=0}^{X-1}\binom{S}{A}p^A(1-p)^{S-A}
$$}
we obtain
$$\sum_{S=0}^{m}\sum_{A=0}^{\lfloor \frac{S}{2} \rfloor}
\left(\binom{S}{A}-\binom{S}{A-1}\right)p^{A}(1-p)^{S-A} =
$$
\[\sum_{S=0}^{m}\binom{S}{\lfloor \frac{S}{2} \rfloor}p^{\lfloor \frac{S}{2} \rfloor}(1-p)^{\lceil \frac{S}{2} \rceil}-
2\epsilon (1+o(1)) \sum_{S=1}^{m}\sum_{A=0}^{\lfloor \frac{S}{2} \rfloor -1}\binom{S}{A}p^A (1-p)^{S-A}.
\]
Claim \ref{cl:sum1} now follows from the following two claims.

\begin{Claim}
\label{cl:sum1a}
\[
\sum_{S=0}^{m}\binom{S}{\lfloor \frac{S}{2} \rfloor}p^{\lfloor \frac{S}{2} \rfloor}(1-p)^{\lceil \frac{S}{2} \rceil}
\geq (1+o(1)) \frac{2 \sqrt{n}}{\alpha}\textnormal{Erf}\left(\frac{\alpha}{\sqrt{2}}\right) .\]
\end{Claim}

\begin{Claim}
\label{cl:sum1b}
\[ 2\epsilon \sum_{S=1}^{m}
\sum_{A=0}^{\lfloor \frac{S}{2} \rfloor -1}\binom{S}{A}p^A (1-p)^{S-A}
\leq
(1+o(1)) \sqrt{n}\left(\alpha-e^{-\frac{\alpha^2}{2}}\sqrt{\frac{2}{\pi}}+\left(\frac{1}{\alpha}-\alpha\right)\textnormal{Erf}\left(\frac{\alpha}{\sqrt{2}}\right)\right) \]
\end{Claim}

\proof [of Claim \ref{cl:sum1a}]
We have 
\[ \binom{S}{\lfloor \frac{S}{2} \rfloor}p^{\lfloor \frac{S}{2} \rfloor}(1-p)^{\lceil \frac{S}{2} \rceil} = (1+o(1)) \frac{\sqrt{2}}{\sqrt{\pi S}} 2^{S} (p(1-p))^{S/2} \]
\begin{equation}
\label{eq:binom}
 = (1+o(1)) \frac{\sqrt{2}}{\sqrt{\pi S}} e^{-\gamma S} 
 \end{equation}
where $\gamma=\frac{-\ln 4p(1-p)}{2}$.
Here, the first equality follows from approximations of binomial coefficients.
We denote $W(n, S) = \frac{\sqrt{2} e^{-\gamma S}}{\sqrt{\pi S}}$.
The identity (\ref{eq:binom}) means that, for every $\delta>0$, there exists $S_0$ 
such that 
\[ \binom{S}{\lfloor \frac{S}{2} \rfloor}p^{\lfloor \frac{S}{2} \rfloor}(1-p)^{\lceil \frac{S}{2} \rceil} \geq (1-\delta) W(n, S) \]
for all $S\geq S_0$.
\comment{By using  
$$
\ln(1 \pm \delta) = \pm \delta \pm O(\delta^2),
$$
we obtain that (\ref{eq:binom}) is equal to
\[ (1+o(1)) \frac{1}{\sqrt{\pi S}} e^{\frac{-S\epsilon^2}{2}+O(S\epsilon^4)} .\]
Since $\epsilon = \frac{\alpha}{\sqrt{n}}$ and $S\leq n$, this is equal to
$(1+o(1)) W(n, S)$ where
\[  W(n, S) = \frac{e^{-\frac{\alpha S}{2n}}}{\sqrt{\pi S}} .\]}
Therefore, we have
$$
\sum_{S=0}^{m}\binom{S}{\lfloor \frac{S}{2} \rfloor}p^{\lfloor \frac{S}{2} \rfloor}(1-p)^{\lceil \frac{S}{2} \rceil} \geq
(1-\delta)  \sum_{S=S_0}^{m} W(n, S) $$
$$\geq (1-\delta) \int_{S_0}^{m}W(n,S)dS$$
\begin{equation}
\label{eq:erf}
=
(1-\delta) \frac{\sqrt{2}}{\gamma} \left( \Erf(\sqrt{\gamma m}) - \Erf(\sqrt{\gamma S_0}) \right)
\end{equation}
Here, the inequality follows from the function $W(n, S)$ being decreasing in $S$.
The last equality follows from $\int \frac{e^{-at}}{\sqrt{t}} dt = 
\frac{\sqrt{\pi} \textnormal{Erf}(\sqrt{at})}{\sqrt{a}}$.
For large $n$, we have
\[ \gamma = \frac{-\ln 4 p(1-p)}{2} = \frac{- \ln (1-\epsilon^2)}{2} =
\frac{- \ln \left(1-\frac{\alpha^2}{n}\right)}{2} =
(1+o(1)) \frac{\alpha^2}{2 n} .\]
We also have $\gamma m= (1+o(1)) \frac{\alpha^2 m}{2 n} = (1+o(1)) \frac{\alpha^2}{2}$
and $\gamma S_0 = (1+o(1)) \frac{\alpha^2 S_0}{2 n} = o(1)$.
Hence, (\ref{eq:erf}) is equal to 
\[ 
(1-\delta) (1+o(1)) \frac{2\sqrt{n}}{\alpha} 
\Erf\left( \frac{\alpha}{\sqrt{2}} \right) .\]
Since this is true for any $\delta>0$, the claim follows. 
\comment{where
$$
\textnormal{Erf}(z)=\frac{2}{\sqrt{\pi}}\int_0^ze^{-t^2}dt.
$$}\qed

\proof [of Claim \ref{cl:sum1b}]
We can interpret $\sum_{A=0}^{\lfloor \frac{S}{2}-1 \rfloor}\binom{S}{A}p^A (1-p)^{S-A}$ 
as follows. Let $X=X_1+\ldots+X_S$ where $X_i$ are independent random variables with
$Pr[X_i=1]=p$ and $Pr[X_i=-1]=1-p$. Then,
\[ \sum_{A=0}^{\lfloor \frac{S}{2} \rfloor -1}\binom{S}{A}p^A (1-p)^{S-A} =
Pr\left[X  \leq \left( \lfloor \frac{S}{2} \rfloor -1 \right) - 
\left( S - \lfloor \frac{S}{2} \rfloor + 1 \right) \right] \leq Pr[X \leq 0] .\]
By Central limit theorem, for large $S$, $Pr[X \leq 0]$ tends to
\[ \frac{1}{2} \Erfc\left(\frac{0-E[X]}{\sqrt{2 D[X]}}\right) = \frac{1}{2} \Erfc\left(\frac{(1-2p) S}{\sqrt{4p S}}\right) \]
where $\Erfc(x)=1-\Erf(x)$.
Since $p=\frac{1}{2}+\frac{\epsilon}{2}$, $\epsilon=\frac{\alpha}{\sqrt{n}}$, this is equal to
\[ \frac{1}{2} \Erfc\left(\frac{\sqrt{S}\epsilon }{\sqrt{2+\epsilon}}\right) 
= (1+o(1)) \frac{1}{2} \Erfc\left(\frac{\sqrt{S}\epsilon }{\sqrt{2}}\right) .\]
Hence,
$$
2\epsilon\sum_{S=1}^{m}\sum_{A=0}^{\lfloor \frac{S}{2} \rfloor-1}\binom{S}{A}p^A (1-p)^{S-A} =
(1+o(1)) \epsilon\sum_{S=1}^{m}\textnormal{Erfc}\left(\frac{\sqrt{S}\epsilon}{\sqrt{2}}\right) 
$$
$$
< (1+o(1)) \frac{\alpha}{\sqrt{n}}\int_{0}^{n}\textnormal{Erfc}
\left(\frac{\sqrt{S}\alpha}{\sqrt{2n}}\right)dS
$$
$$
= (1+o(1)) \sqrt{n}\left(\alpha-e^{-\frac{\alpha^2}{2}}\sqrt{\frac{2}{\pi}}+\left(\frac{1}{\alpha}-\alpha\right)\textnormal{Erf}\left(\frac{\alpha}{\sqrt{2}}\right)\right).
$$
Here, the inequality follows from $\Erfc$ being decreasing and $m<n$ and
the last equality follows from 
\[ \int \Erfc (a\sqrt{t}) dt = 
\frac{\Erf(a \sqrt{t})}{2 a^2} - \frac{\sqrt{t} e^{-a^2 t}}{\sqrt{\pi} a} + t (1-\Erf(a\sqrt{t})) .\]
\qed

\subsection{Proof of Claim \ref{cl:sum2}}

Let $m=\lfloor \frac{n-1}{2} \rfloor$.
Because of (\ref{eq:middle}), we have
$$
\sum_{S=m}^{2 m - 2}\sum_{A=0}^{S-m}
\left(\binom{S}{A}-\binom{S}{A-1}\right)p^{A}(1-p)^{S-A}
$$
$$
= \sum_{S=m}^{2 m -2}\binom{S}{m}
p^{S-m}(1-p)^{m}-\frac{2\epsilon}{1-\epsilon}
\sum_{S=m}^{2 m -2}\sum_{A=0}^{S-m-1}
\binom{S}{A}p^{A}(1-p)^{S-A}
$$
$$
< \sum_{S=m}^{2 m -2}
\binom{S}{m}2^{-S}
$$
where the last inequality follows from 
$$
p^{A}(1-p)^{S-A}\leq 2^{-S}
$$
that holds if $2A\leq S$.
Claim \ref{cl:sum2} now follows from 

\begin{Claim} 
Let $m$ be a positive integer. Then,
$$
\sum_{S=m}^{2m}\binom{S}{m}2^{-S}=1.
$$
\end{Claim}

\proof
We give a combinatorial proof of this equality.

Consider a random process  where we have two boxes each containing $m+1$ pebbles. At each 
step we choose one of the two boxes with equal probability $\frac{1}{2}$ and remove one 
pebble from this box. The process finishes when we remove the last pebble from one of the two 
boxes.

We calculate
$$
\textnormal{Pr}\left[\textnormal{The process finishes after }k\textnormal{ steps}\right]=
$$
$$
2\textnormal{Pr}\left[\textnormal{The process finishes after }k\textnormal{ steps and the first box is empty}\right]
$$

To empty the first box we need to choose the first box in $m$ out of $k-1$ steps which happens with a probability of $\binom{k-1}{m}\frac{1}{2^{k-1}}$ and in the last step we need to choose the first box which happens with a probability of ${1 \over 2}$. Thus
$$
\textnormal{Pr}\left[\textnormal{The process finishes after }k\textnormal{ steps}\right]=2\binom{k-1}{m}\frac{1}{2^{k-1}}\frac{1}{2}=\binom{k-1}{m}\frac{1}{2^{k-1}}.
$$

The number of steps $k$ can be any value from  $m+1$ to $2m+1$ including. Because the process must end after some number of steps we obtain
$$
\sum_{k=m+1}^{2m+1}\binom{k-1}{m}\frac{1}{2^{k-1}}=1.
$$
Claim \ref{cl:sum2} follows by substituting $k=S+1$.
\qed

\end{appendix}
\end{document}